\documentclass{aa}  
\usepackage{graphicx}
\usepackage{txfonts}
\usepackage{natbib}
\usepackage{color}
\usepackage{hyperref}
\hypersetup{colorlinks=true,allcolors=[rgb]{0,0,0.8}}
\makeatletter
\renewcommand*\aa@pageof{, page \thepage{} of \pageref*{LastPage}}

\newcommand{\rjup}{$R_{\rm Jup}$}
\newcommand{\mjup}{$M_{\rm Jup}$}

\usepackage[nolist]{acronym}

\newacro{bic}[BIC]{Bayesian Information Criterion}
\newacro{sde}[SDE]{Signal Detection Efficiency}
\newacro{tls}[TLS]{Transit Least Squares}

\makeatother

\begin{document} 

   \title{A search for transiting companions in the J1407 (V1400 Cen) system\thanks{Processed photometric data are available in electronic form at the CDS via anonymous ftp to \url{cdsarc.u-strasbg.fr} (130.79.128.5) or via \url{http://cdsweb.u-strasbg.fr/cgi-bin/qcat?J/A+A/}}}

   \author{S.~Barmentloo\inst{1}\thanks{The first two authors contributed equally to this work.}
          \and
          C.~Dik\inst{1}$^{**}$
          \and
          M.~A.~Kenworthy\inst{1}
          \and
          E.~E.~Mamajek\inst{2,3}
          \and
          F.-J.~Hambsch\inst{4,5}
          \and
          D.~E.~Reichart\inst{6}
          \and
          J.~E.~Rodriguez\inst{7}
          \and
          D.~M.~van Dam\inst{1}
          }

   \institute{Leiden Observatory, University of Leiden,
              PO Box 9513, 2300 RA Leiden, The Netherlands\\
             \email{kenworthy@strw.leidenuniv.nl}
        \and
Jet Propulsion Laboratory, California Institute of Technology, 4800 Oak Grove Drive, M/S 321-100, Pasadena, CA, 91109, USA
\and 
Department of Physics \& Astronomy, University of Rochester, Rochester, NY 14627, USA
\and
American Association of Variable Star Observers (AAVSO), 49 Bay State Rd., Cambridge, MA 02138, USA
\and
Vereniging Voor Sterrenkunde (VVS), Oostmeers 122 C, 8000 Brugge, Belgium
\and
Department of Physics and Astronomy, University of North Carolina at Chapel Hill, Campus Box 3255, Chapel Hill, NC 27599, USA
\and
Department of Physics and Astronomy, Michigan State University, East Lansing, MI 48824, USA
             }

   \date{Received 9 March 2021; accepted 17 June 2021}

  \abstract
   {In 2007, the young star 1SWASP J140747.93-394542.6 (V1400 Cen) underwent a complex series of deep eclipses over 56 days.
   This was attributed to the transit of a ring system filling a large fraction of the Hill sphere of an unseen substellar companion.
   Subsequent photometric monitoring has not found any other deep transits from this candidate ring system, but if there are more substellar companions and they are coplanar with the potential ring system, there is a chance that they will transit the star as well.
   This young star is active and the light curves show a 5\% modulation in amplitude with a dominant rotation period of 3.2 days due to star spots rotating in and out of view.
}
   {We model and remove the rotational modulation of the J1407 light curve and search for additional transit signatures of substellar companions orbiting around J1407.}
   %
   {We combine the photometry of J1407 from several observatories, spanning a 19 year baseline.
   We remove the rotational modulation by modeling the variability as a periodic signal, whose periodicity changes slowly with time over several years due to the activity cycle of the star. 
   A Transit Least Squares (TLS) analysis searches for any periodic transiting signals within the cleaned light curve.
   }
   {We identify an activity cycle of J1407 with a period of 5.4\,yr.
   A Transit Least Squares search does not find any plausible periodic eclipses in the light curve, from 1.2\% amplitude at 5 days up to 1.9\% at 20 days.
   This sensitivity is confirmed by injecting artificial transits into the light curve and determining the recovery fraction as a function of transit depth and orbital period.}
   {J1407 is confirmed as a young active star with an activity cycle consistent with a rapidly rotating solar mass star.
   With the rotational modulation removed, the TLS analysis reaches down to planetary mass radii for young exoplanets, ruling out transiting companions with radii larger than about 1 \rjup{}.}


   \maketitle

\section{Introduction}

Ring systems are a ubiquitous feature in planetary systems - all the gas giants in the Solar System have ring systems around them of varying optical depths \citep[see e.g.][]{2013pss3.book..309T,2018haex.bookE..54C}, and ring systems have been detected around minor planets \citep[e.g. Chariklo;][]{2014Natur.508...72B}, so it is reasonable that exoplanets and substellar objects host ring systems too.
Long period eclipsing binary star systems, where one star is surrounded by an extended dark disk-like structure that periodically eclipses the other component, have already been observed, such as EE Cep \citep{10.1046/j.1365-8711.1999.02257.x}, $\epsilon$ Aurigae \citep{2002ASPC..279..121G} and TYC 2505-672-1, with a companion period of 69 years \citep{2016A&A...588A..90L,2016AJ....151..123R}.
A large ring-like structure around a substellar companion was proposed to explain observations from 2007 from the J1407 system \citep{2012AJ....143...72M}.
1SWASP J140747.93-394542.6 (V1400 Cen; hereafter called `J1407'), is a young, pre-main sequence star in the Sco-Cen OB association \citep{2012AJ....143...72M} with spectral type K5 IV(e) Li and is similar in size and mass to the Sun .
In 2007, it displayed a complex symmetric dimming pattern of up to $\sim$3 magnitudes during a 56 day eclipse.
This has been attributed to the transit of a substellar companion (called ``J1407\,b'') with a mass of 60 to 100\mjup{} \citep{Rieder_2016} surrounded by an exoring system consisting of at least 37 rings and extending out to 0.6\,au in radius \citep{Kenworthy_2015}.
For these rings to show detectable transit signatures, they must be significantly misaligned with respect to the orbital plane of J1407\,b \citep{Zanezzi}.
This potential ring system would be considerably larger than the ring system of Saturn, which is located within the planet's tidal disruption radius.
The proposed rings around J1407\,b would even cover a significant fraction of the companion's Hill sphere, and would not expected to be stable over gigayear timescales.
If the candidate ringed companion is in a bound orbit around the star, this orbit must be moderately eccentric in order for no other eclipses to have been detected to date \citep{Kenworthy15}, raising the possibility that there might be a second as yet undetected companion in the system that causes the implied orbital eccentricity for J1407\,b.
Radial velocity measurements are overwhelmed by the chromospheric noise of the star and do not place strong constraints on other substellar companions \citep{Kenworthy15}.
The transit of J1407 suggests that its orbital plane has a high inclination to our line of sight - if there are other planets inside the orbit of J1407\,b, their orbits may well be coplanar with J1407\,b and there is a high chance that these companions may transit J1407.

As J1407 is a young ($\sim$16 Myr), active star \citep{Kenworthy15}, and its brightness changes on timescales of days (rotational modulation) to years (activity cycles).
The behaviour of stellar magnetic fields is closely linked to the number of star spots we can observe.
As these star spots are cooler than their surroundings, their presence can noticeably affect the luminosity of the star.
A distant observer would see this as a periodic variability in the light curve of the star as they rotate in and out of the line of sight.
Although the exact pattern of this variability is more complex, on the time interval of the observations it can be approximated as a combination of a long-term linear trend and a short-term sinusoidal trend with a varying period of modulation.
For J1407, these spots cause a $\sim$0.1 mag sinusoidal variability with a periodicity of $\sim$3.2 days, corresponding to the rotation period of the star.
This variability complicates the search for transit signals.

In this paper, we look for transiting exoplanets in the J1407 system by determining and then removing a model of stellar activity from the photometry of several ground-based telescopes, and from photometry from Transiting Exoplanet Survey Satellite (TESS) \cite{2015JATIS...1a4003R}.
By analysing the combined and corrected light curve of different data sets we put constraints on the size and period of possible transiting substellar companions. 
The data is presented in Section~\ref{sec:data}, and the methodology for manipulating and analysing the data is in Section~\ref{sec:analysis}.
Results on the long term starspot cycle as well as constraints on additional transiting exoplanets can be found in Section~\ref{sec:result}.
An interpretation of these results are found in Section~\ref{sec:discuss} followed by the conclusions in Section~\ref{sec:conclusion}.
    
\section{Data}\label{sec:data}

\subsection{Ground-based Telescopes}

\begin{figure}[htb]
    \centering
    \includegraphics[width = \hsize, height = 0.5\hsize]{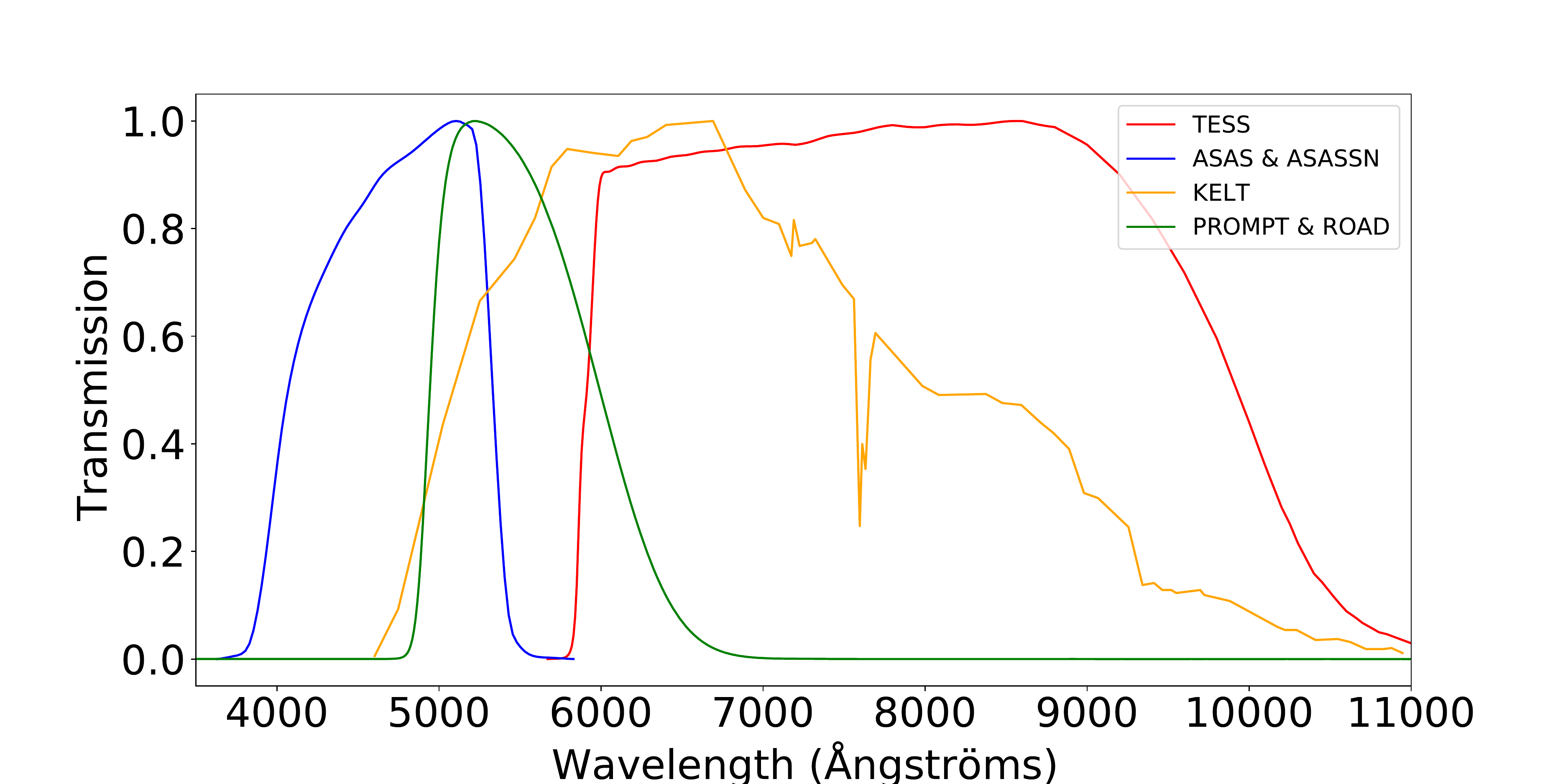}
    \caption{The response curves for all the instruments in this paper, normalised at their peak transmission wavelength.}
    \label{fig:response}
\end{figure}

The photometric data is from five different ground-based telescopes, resulting in an observational baseline of $\sim$19 years (see Table~\ref{tab:tel}).
The All Sky Automatic Survey (ASAS) monitors around 10 million stars up to magnitude 14 in the $I$ and $V$ bands using observing stations in Hawaii and Chile \citep{1997AcA....47..467P}.
The All Sky Automated Survey for Supernovae (ASAS-SN) surveys the entire sky up to stars with a $V$-magnitude of 17, looking for signals of variability with observing stations all around the globe in for example Hawaii, Chile, South-Africa and China \citep{2017PASP..129j4502K}. 
The final three sets of observations are taken from \citet{2018A&A...619A.157M}.
The first set of these three contains observations from the Kilodegree Extremely Little Telescope \citep[KELT;][]{2007PASP..119..923P,Pepper12} of J1407 between 2010 and 2015. 
These observations were made using the KELT-south telescope, located in South Africa, which surveys a field of 26$^{\circ}$ by 26$^{\circ}$ in the southern sky, searching for transiting Hot Jupiters \citep{Pepper12}.
The Panchromatic Robotic Optical Monitoring and Polarimetry Telescopes (PROMPT; \cite{2005NCimC..28..767R}) is a network of telescopes in North Carolina and Chile used to detect gamma ray burst afterglows.
Observations of J1407 in the Johnson V band by the PROMPT-4 and PROMPT-5 telescopes in Chile are described in \citet{2018A&A...619A.157M}.
Johnson V band data from the 40-cm Remote Observatory Atacama Desert \citep[ROAD;][]{2012JAVSO..40.1003H} are included, spanning from mid-2012 up to 2020. 
The response functions for each instrument (including the TESS telescope) are shown in Figure~\ref{fig:response}. 

\begin{figure*}[htb]
    \centering
    \includegraphics[width = \hsize]{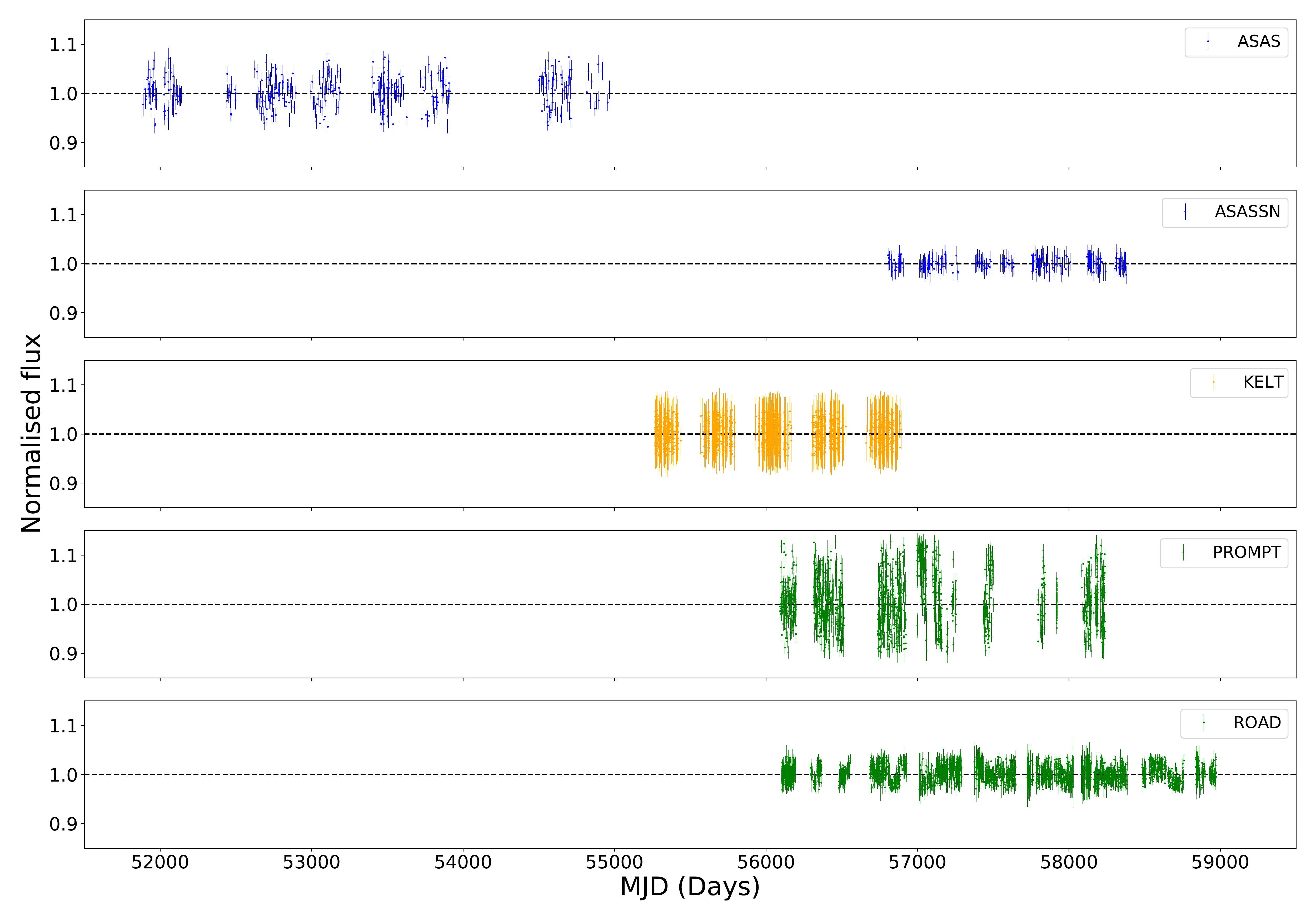}
    \caption{Light curves for the five ground-based telescopes after correcting for both a downward linear trend and the dominant periodicities as described in Section~\ref{dominant_removal}. The 5\% lowest and 5\% highest of the flux values for each separate telescope were removed.}
    \label{fig:groundbased}
\end{figure*}

We remove all measurements with photometric r.m.s. error $> 5\%$, and are left with 10941 data points.
The standard deviation of the normalised flux is 0.041 (equivalent to the transit depth of an object with radius 1.9\,\rjup{}, for this specific star).
The images are taken at an irregular cadence, averaging about one image per day.
As all the long cadence time series are from ground-based telescopes, the data contains gaps due to both the diurnal and annual observing windows.
The combined photometry of the telescopes is shown in Figure~\ref{fig:groundbased}.

\begin{table*}[ht]
\caption{Data coverage of J1407 from ground-based telescopes.}\label{tab:tel}
\centering
\begin{tabular}{l c c c c}
\hline\hline
Telescope & First Obs. (MJD) & Last Obs. (MJD) & No. of Obs. & Reference \\ \hline
ASAS & 51887 & 54966 & 429 & 1 \\ 
ASAS-SN & 56805 & 58377 & 248 & 2 \\ 
KELT & 55268 & 56893 & 4699 & 3,4 \\ 
PROMPT & 56092 & 58236 & 1980 & 5 \\ 
ROAD & 56106 & 58971 & 3585 & 6 \\ \hline
\end{tabular}%
\tablebib{
(1)~\citet{1997AcA....47..467P}; (2) \citet{2017PASP..129j4502K}; (3) \citet{2007PASP..119..923P}; (4) \citet{Pepper12}; (5) \citet{2005NCimC..28..767R}; (6) \cite{2012JAVSO..40.1003H}.
}
\end{table*}

\subsection{TESS}

\begin{figure}[htb]
    \centering
    \includegraphics[width = \hsize]{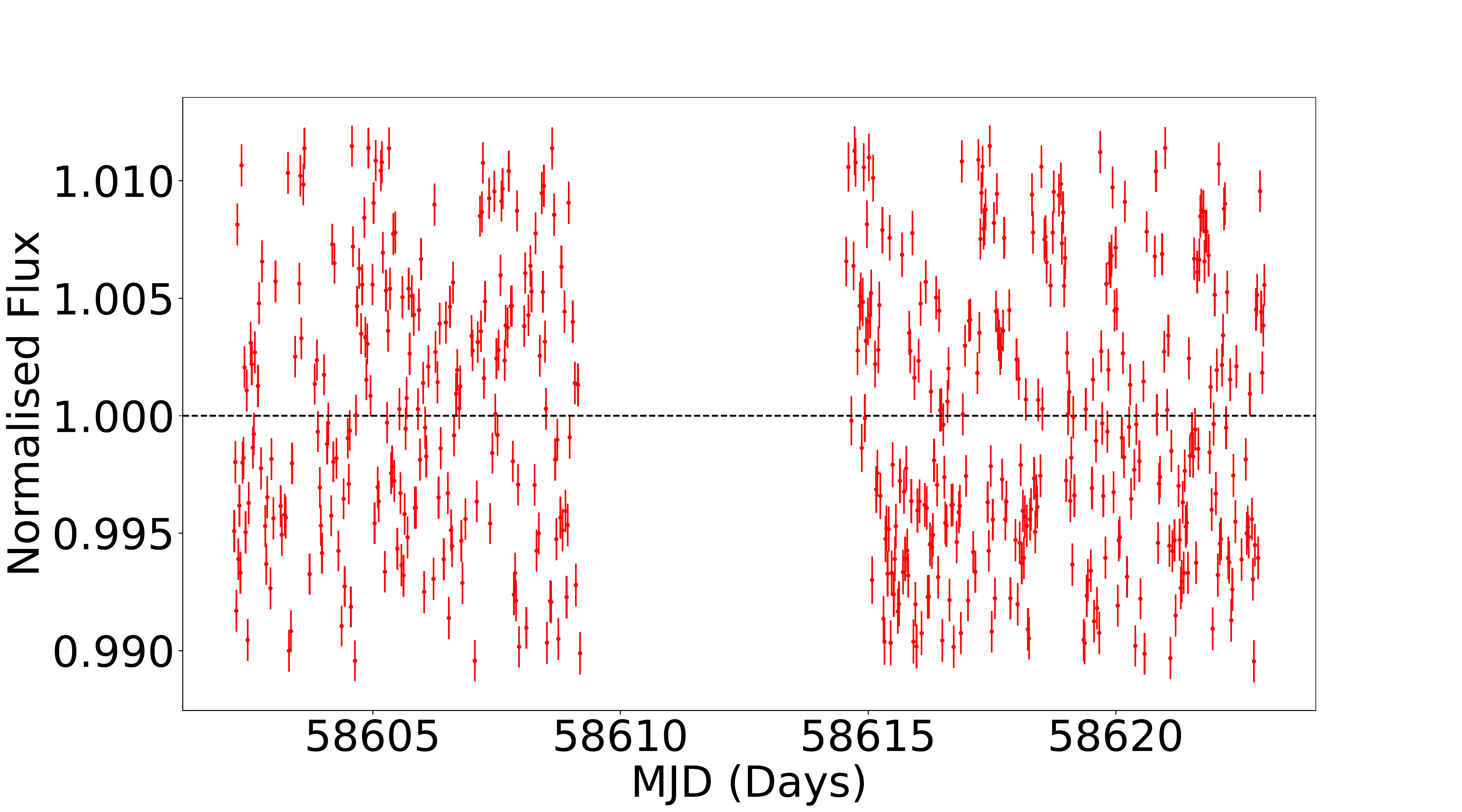}
    \caption{Light curve of the TESS data after removing the dominant periodicities via the routine described in Section~\ref{dominant_removal}. The 5\% lowest and 5\% highest of the flux values were removed.}
    \label{fig:tess}
\end{figure}

The Transiting Exoplanet Survey Satellite \citep[TESS; ][]{2015JATIS...1a4003R} is an MIT-led NASA mission to search for transiting exoplanets around bright stars from the Galactic poles down to the Galactic plane.
TESS has observed 26 segments of the sky with a 27.4 day observational period per segment \cite{2015JATIS...1a4003R}, making it sensitive to exoplanets with orbital periods shorter than 13 days. 
While observing a segment, TESS returns full frame images in a photometric bandpass between 600 to 1000 nm similar to Cousins $I$-band at a 30-minute cadence. 
For stars with TESS magnitude 9-15, precisions on the order of 50 p.p.m are acquired.
J1407 was observed by TESS in Sector 11 from 22 April 2019 to 21 May 2019.
The TESS light curves of J1407 were extracted from the MAST archive using the {\tt eleanor} \citep{2019PASP..131i4502F} software package.
The data for J1407 contains 744 observations over 23 days (a few days are missing due to light scattered from Earth contaminating the data) and the resultant light curve is shown in Figure~\ref{fig:tess}.
The standard deviation is about 0.014 for the normalised flux  (equivalent to the transit depth of an object with radius 1.1\,\rjup{} for this specific star). 

\section{Analysis}\label{sec:analysis}

\subsection{Removal of long term trends}
\label{sec:stellar_variability}

In order to obtain the best sensitivity for planetary transits, the light curve needs to be corrected for the star's variability.
First, a long term linear trend was removed by fitting a linear function to the light curve\footnote{The entire code used for the analysis and creating the figures in this paper is available at \url{https://github.com/StanBarmentloo/J1407_transit_search_activity}}.
The fit was applied and removed for each of the ground-based data sets separately.
The nature of this long-term trend will be discussed in more detail in a separate paper (Barmentloo et al. in preparation), but we note that the effect occurs in the ROAD, ASAS and ASAS-SN datasets, while not showing up in the KELT and PROMPT data.
The effect occurs most strongly (both in gradient and in Signal to Noise ratio) in the ROAD data, where it has a gradient of  $-1.06 \pm 0.02 \%$ per year from 2012 up to 2020.
The gradient seen in the ASAS dataset, which covers 2001 up to 2009 and is therefore independent from the ROAD dataset, is similar at $-1.07 \pm 0.12 \%$ per year.
A search in the ASAS data for stars around J1407 found a similar trend in a non negligible number of stars, with preferentially negative gradients distributed around an average of about $-1\%$ per year.
Searching for the trend in ASAS in an equally sized area in a random part of the sky found no such trend at all, which suggests that this effect is of astrophysical origin.

\subsection{Removing the rotational modulation}
\label{dominant_removal}

\begin{figure}[htb]
    \centering
    \makebox[\hsize][c]{\includegraphics[width=\columnwidth]{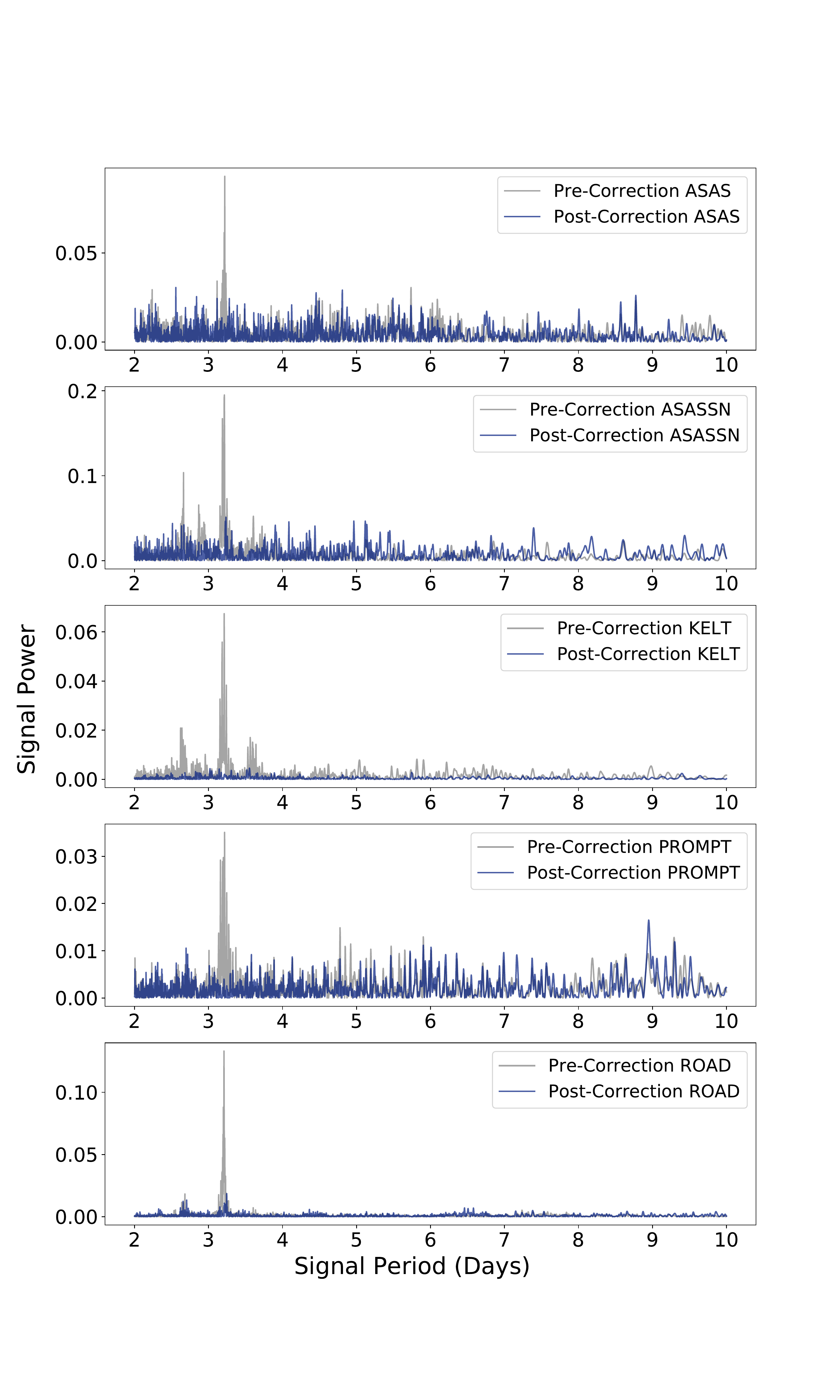}}
    \caption{Comparison of the Lomb-Scargle periodograms before and after applying the routine from Section~\ref{dominant_removal} for each of the ground-based telescopes}
    \label{fig:dominant_removal}
\end{figure}

\begin{figure}[htb]
    \centering
    \includegraphics[width = \hsize]{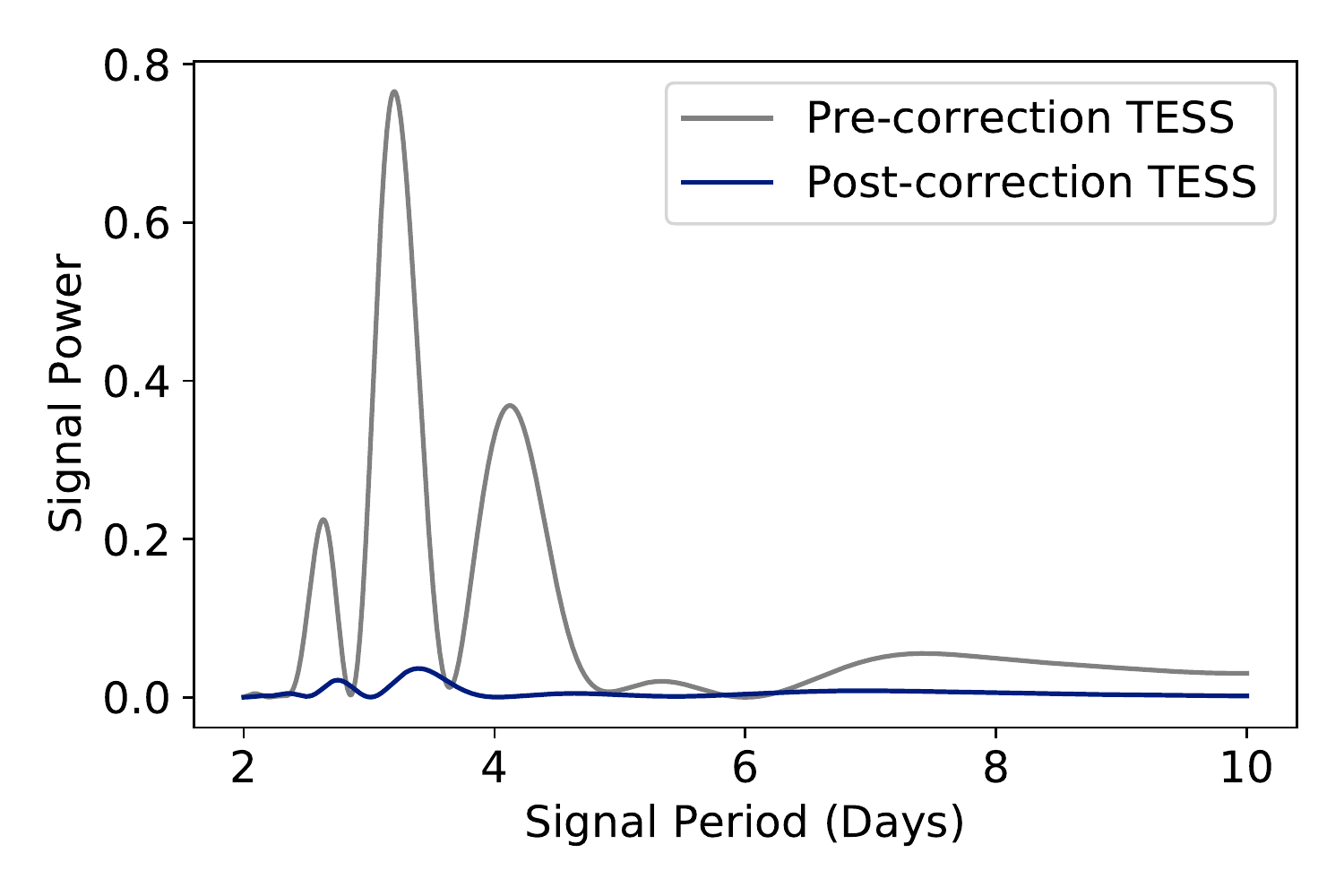}
    \caption{Same as Figure~\ref{fig:dominant_removal}, but for the TESS data.}
    \label{fig:tess_periodogram}
\end{figure}

To further increase the sensitivity of the TLS search (see Section~\ref{sec:TLS}, strong periodic signals in the lightcurve should be suppressed. 
To this end, each separate ground-based telescope data set was divided into segments.
A new segment was started either when there was a 10 day gap between photometric points or when the segment would be longer than 500 days.
For each of these segments, a Lomb-Scargle (LS) periodogram was calculated. The peak signal (most often the 3.2 day signal, but sometimes also with another periodicity) was noted. 
Next, the segment was time folded over the periodicity of exactly this peak signal and to this a superposition of sines and cosines with different frequencies was fitted, given by the formula:

\begin{equation}
    F = c + \sum_{n=1}^{2} a_{n} \sin (n\phi) + b_{n} \cos (n\phi)
\end{equation}

where \textit{F} is the fitted flux, \textit{a$_{n}$} and \textit{b$_{n}$} are the best-fit amplitudes and \textit{c} is the best-fit offset.
The best fit curve was then subtracted from the time folded flux in each segment.
As most segments still had a relatively strong periodic signal remaining, the process of determining the dominant segment signal and removing a best fit curve was performed twice for each segment. 
The substantial decrease in power of periodic signals caused by this routine is shown for each telescope in Figure~\ref{fig:dominant_removal}.
Finally, the now doubly corrected segments were stuck back together and the 5\% lowest and 5\% highest of the flux values were removed for each separate telescope. 
We removed these outliers as they, due to their depth, could not possibly have been part of a transit.
The TLS algorithm would however try to take these outliers into account when fitting a transit, causing it to possibly miss injected transits.
The light curves of the five separate telescopes were combined into one after the above routine was performed. The resulting lightcurve is shown in Figure~\ref{fig:groundbased} .

For the TESS data the same approach was used, with the only difference being the segment length: as the TESS data only spans ~25 days, the entire dataset was considered as a single segment. 
Again, the routine was performed twice to also remove signals less dominant than the 3.2 day signal. 
The resulting change in periodogram can be seen in Figure~\ref{fig:tess_periodogram}.

\begin{figure}[htb]
    \centering
    \makebox[\hsize][c]{\includegraphics[width=\hsize]{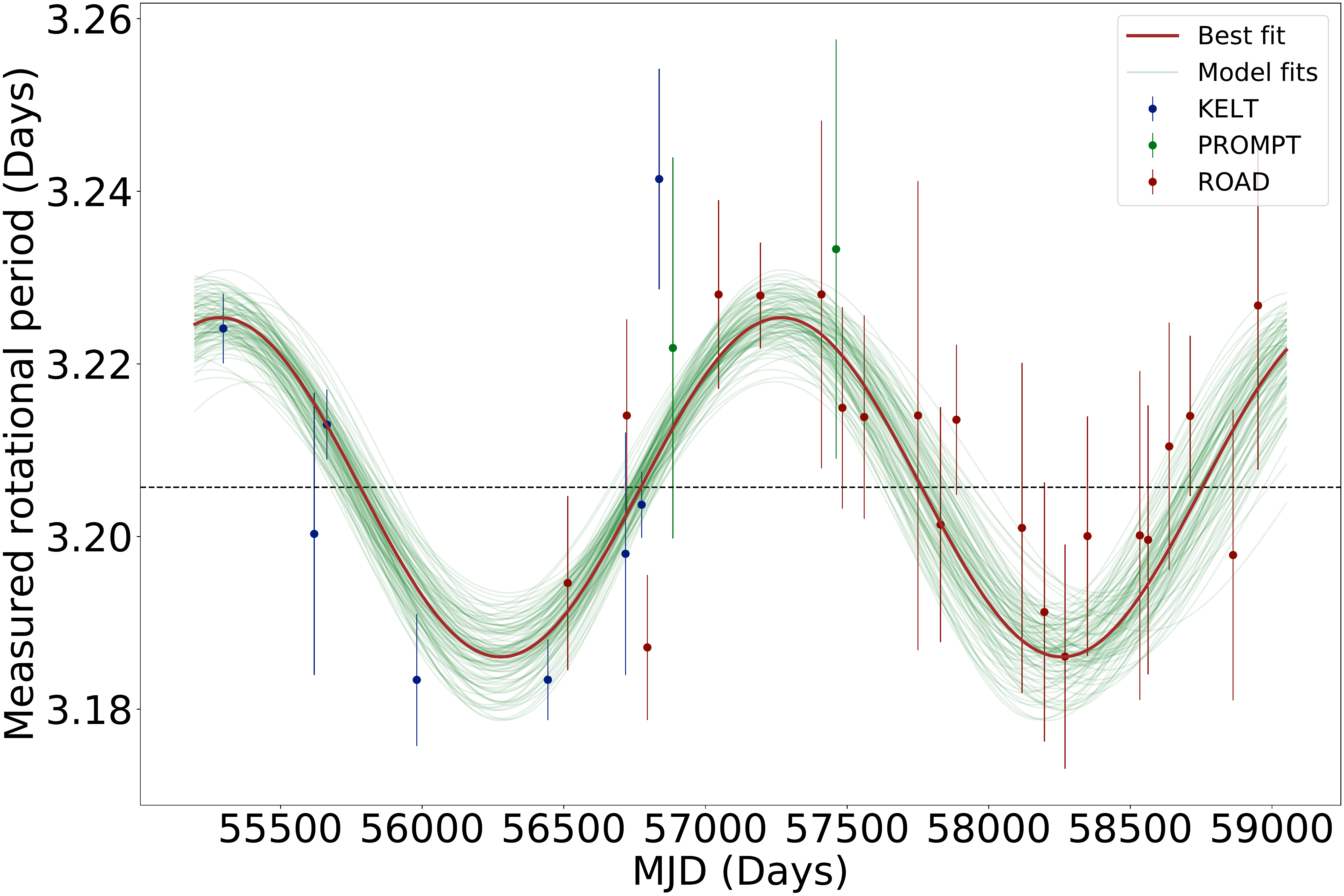}}
    \caption{Fitting of the long term activity of J1407. The brown-red curve indicates the best fitting sinusoid. The black dotted line shows the (weighted) average rotation period. }
    \label{fig:activity_cycle}
\end{figure}

We looked for long term variations in the $\sim$3.2 day signal in the ground-based data.
The light curves from each ground-based telescope (in the form that they have after the correction in Section~\ref{sec:stellar_variability}, but before the corrections in Section~\ref{dominant_removal}) were individually divided at points where there were large gaps in the observations into segments of data.
These segments were then divided into segments of 75 days duration, which we determined was the optimal value minimising the determined rotational period error (for shorter segment size) and preventing the under-sampling of the activity cycle (for longer segment durations). 
If the segment itself spanned less than 75 days, it was simply considered as one segment.
For a segment to be considered, there were two further requirements: firstly, the segment should have an average sampling of at least 2.5 data points per 3.2 days to adhere to the Nyquist sampling theorem (this criterion was never met by ASAS and ASAS-SN).
Secondly, we required the Lomb-Scargle periodogram from 2 to 5 days to have its peak value between 3.1 and 3.3 days, to avoid sampling a segment with a different dominant period than the 3.2 period.
A Lomb-Scargle periodogram was then calculated on each valid segment and the highest peak in the periodogram between 3.1 and 3.3 days was taken to be the mean spot rotational period during the midpoint in time $t$ of that segment, called $P_{rot}(t)$.
The errors were determined using a Bootstrap technique by generating data with the same cadence and standard deviation as the specific segment and adding to this a sine with typical amplitude and 3.2 day periodicity.
The retrieved periodicities from multiple runs of the LS periodogram then form a Gaussian distribution, of which the standard deviation was taken to be the error on the determined period for that segment.
These measured periods are shown in Figure~\ref{fig:activity_cycle} and tabulated in Table~\ref{tab:dat}.

The {\tt Python} package {\tt emcee} \citep{emcee} was used to determine the activity cycle period by fitting a model to the data.
Leaving out unknown asymmetries in the stellar activity, the activity cycle is periodic and on average well described by a sinusoid \citep{2020A&A...638A..69W}.
We modeled the activity cycle $P_{activity}$ as:

$$P_{rot}(t) = P_{meanrot}+a\sin (2\pi t/P_{activity}+\phi) $$

where $P_{meanrot}$ is the mean spot rotational period, $a$ is the amplitude of the variation of the spot rotational period, and $\phi$ determines the phase of the sinusoidal fit.
An initial fit was carried out by the {\tt lmfit} \citep{lmfit} package in {\tt Python}, and then these values of $P_{meanrot}, a, P_{activity}, \phi$ were used as initialisation points for the walkers in the {\tt emcee} routine.

\begin{table*}[ht]
\caption{All data points as plotted in Figure~\ref{fig:activity_cycle}. The data points are ordered per telescope by first observation.}
\label{tab:dat}
\centering
\begin{tabular}{l c c c c}
\hline\hline
First Obs. & Last Obs. & Period & Error & Telescope \\
(MJD) & (MJD) & (Days) & (Days) & \\ \hline
55267 & 55338 & 3.224 & 0.004 & KELT \\
55569 & 55622 & 3.200 & 0.016 \\
55645 & 55718 & 3.213 & 0.004 \\
55932 & 56004 & 3.183 & 0.008 \\
56422 & 56485 & 3.183 & 0.005 \\
56669 & 56731 & 3.198 & 0.014 \\
56743 & 56808 & 3.204 & 0.004 \\
56811 & 56884 & 3.241 & 0.013 \\
\hline
57435 & 57499 & 3.233 & 0.024 & PROMPT \\
56855 & 56924 & 3.222 & 0.022 \\
\hline
56480 & 56550 & 3.195 & 0.010 & ROAD \\
56686 & 56760 & 3.214 & 0.011 \\
56761 & 56835 & 3.187 & 0.008 \\
57011 & 57085 & 3.228 & 0.011 \\
57162 & 57235 & 3.228 & 0.006 \\
57374 & 57448 & 3.228 & 0.020 \\
57449 & 57524 & 3.215 & 0.012 \\
57525 & 57599 & 3.214 & 0.012 \\
57724 & 57798 & 3.214 & 0.027 \\
57783 & 57858 & 3.201 & 0.014 \\
57859 & 57932 & 3.214 & 0.009 \\
58082 & 58146 & 3.201 & 0.019 \\
58164 & 58232 & 3.191 & 0.015 \\
58233 & 58307 & 3.186 & 0.013 \\
58308 & 58381 & 3.200 & 0.014 \\
58483 & 58557 & 3.200 & 0.019 \\
58528 & 58602 & 3.200 & 0.016 \\
58604 & 58675 & 3.210 & 0.014 \\
58679 & 58752 & 3.214 & 0.009 \\
58837 & 58899 & 3.198 & 0.017 \\
58927 & 58971 & 3.227 & 0.019 \\
\hline
\end{tabular}%
\end{table*}

\subsection{TLS search for transits}

\label{sec:TLS}

An optimized detection algorithm to search for transits is the \ac{tls} algorithm \citep{Hippke_2019}, which improves on the classical Box Least Squares transit searching algorithm \citep{Kov_cs_2002}.
\ac{tls} computes the \ac{sde} of each signal, which can be used to determine the uncertainty on the detection or to constrain the parameters that the given data would be sensitive to.
For \ac{tls}, the \ac{sde} threshold for a false positive rate of 1\% in simulated white noise data is \ac{sde}\, =\, 7.
There is discussion as to what \ac{sde} value should be considered a confirmed transit.
In the literature \citep{Hippke_2019}, detection thresholds from between \ac{sde}\, =\, 6 and \ac{sde}\, =\, 10 can be found.
During the transit-injection retrieval experiment described below, we adopt a detection threshold value of 6, as we know the orbital parameters of the injected transits a priori.
We insert artificial transits into the ground-based telescope data and determine the retrieval fraction for different values of the orbital period and planetary radius.
For the creation of fake transit signals, the {\tt BATMAN} package \citep{Kreidberg_2015} was used.
The parameters used for the star and exoplanet's orbit are given in Table~\ref{orbitalparam}.
For the transits created by {\tt BATMAN}, a non-linear, four parameter limb darkening model was used with the specific limb darkening coefficients for J1407 approximated as a $T_{eff}$ = 4500\,K, $\log(g)\, =\, 4.0 \, m\, s^{-2}$ surface gravity star, and the corresponding values from \cite{2011yCat..35290075C} were used.

\begin{table}[ht]
\caption{Parameters for the injected transits.}
\label{orbitalparam}
\centering
\begin{tabular}{ l c c}
\hline \hline
Orbital Parameter & Value & References \\ \hline
Mass star & 0.95 M$_{\odot}$ & 1 \\
Radius star & 0.96 R$_{\odot}$ & 2 \\
Orbital Inclination & $90^{\circ}$ & \\
Eccentricity & 0 &  \\
Longitude of Periastron & $90^{\circ}$ & \\ \hline
\end{tabular}%
\tablebib{
(1)~\citet{2018A&A...619A.157M}; (2) \citet{2012AJ....143...72M}.
}
\end{table}

The planetary radius was varied in steps of 0.15\,\rjup{} from 0.5 up to 2.0\,\rjup{}.
The orbital period was varied in 100 logarithmically spaced steps from 3 to 40 days. 
For every combination of planetary radius and orbital period, ten trial runs were done. 
For each try an arbitrary time of inferior conjunction (offset) between 0 and the inserted orbital period was picked from a uniform distribution.
This process was performed twice: once with artificial transits inserted into the activity removed light curve and once with the transits inserted in an artificial light curve that included white (Gaussian) noise, with identical observation times and flux errors as the real data but with randomised flux values.

\section{Results}\label{sec:result}

\subsection{Characterization of the Rotation Period Variability}\label{sec:perper}

The exact periodicity of starspots varies over a timescale of years, which we attribute to starspots migrating to and from the equator.
Due to the differential rotation of the star, there are different rotational periods at each latitude, analogous to the 11 year Schwabe cycle of the Sun \citep{Hathaway_2015}. 
In Figure~\ref{fig:activity_cycle}, after removing best fit period values at the edge of the period grid (less than 3.1 or greater than 3.3 days, indicating a failed fit), the most likely periods were plotted at the midpoints of the intervals used for the calculation and colour-coded for the respective source.
A sinusoidal pattern is visible in Figure~\ref{fig:activity_cycle}.
Further confirmation is provided when determining the difference in the \ac{bic} between a linear fit and the best sine fit.
We determine the \ac{bic} values using the weighted least squares version of the Gaussian case. 
Despite the penalty of having four free parameters compared to the two free parameters for a linear fit, the BIC value for the sine fit is about 32 times lower than that for a linear fit (9 and 41 respectively). 
This delta BIC confirms the perceived sinusoidal pattern. 

The best-fitting value for the period cycle of J1407 is found to be $P_{cyc} = 1981_{-83}^{+94}$ days, where the calculated rotation period oscillates around the median $P_{rot} = 3.206\pm0.002$ days with an amplitude $a$ of $0.020\pm 0.003$ days.
We note that the logarithm of the ratio of the periods is $\log P_{rot} / P_{cyc} = -2.79$, and that this is at least consistent with other young K-type stars \citep{cycle_periods_saar_1999}.
J1407 is a young star and has X-ray emission consistent with saturation \citep[$\log L_X/L_{BOL}=-3.4$;][]{2012AJ....143...72M} where saturation is typically seen around $-3.2\pm 0.2$, and this is expected for rotation periods less than about 3.5 days for K5 stars \citep{Pizzolato03}.
The star is therefore in a regime where the magnetic activity is no longer following the magnetic activity/rotational period relation, and so interpreting the periodic variation of the rotational period in terms of the dynamo model of magnetic activity should be treated with caution.

\subsection{Transit Search}

\begin{figure}[htb]
    \centering
    \includegraphics[width = \hsize]{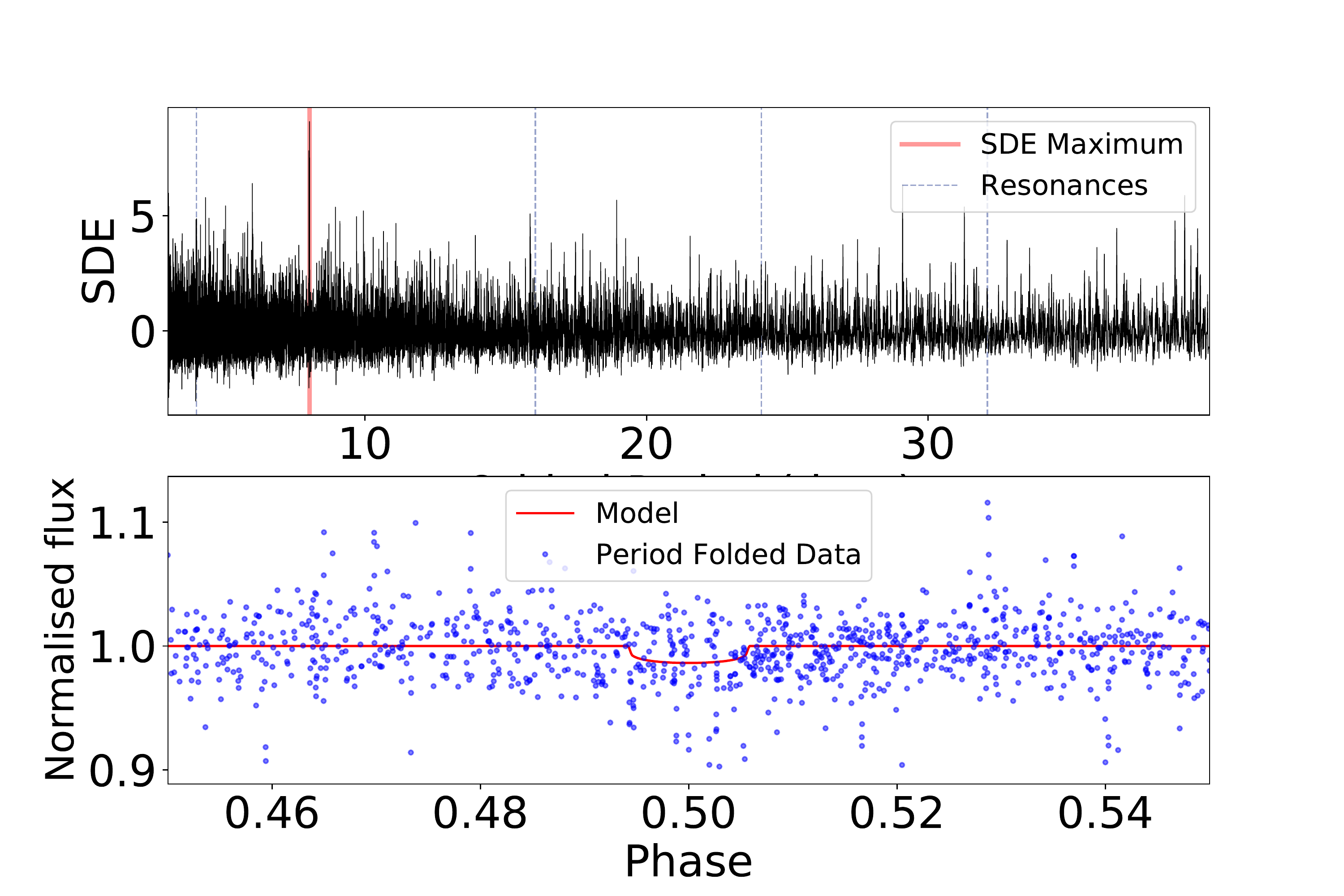}
    \caption{Upper plot: \ac{tls} spectrum for the ground-based data. The \ac{tls} maximum has a value of 9.09 and is located at an orbital period of 8.03 days. Lower plot: ground-based data folded over the most likely orbital period. The fitted model for the transit has a depth of 1.4\% (1.1 \rjup{}) and a duration of 0.093 days.}
    \label{fig:groundtls}
\end{figure}

A \ac{tls} planet search was performed on the data from Figure~\ref{fig:groundbased}.
Values between 3 and 40 days for the orbital period were considered.
The resulting \ac{sde} spectrum is given in the upper panel of Figure~\ref{fig:groundtls} and shows no significant periodicities.
The highest \ac{tls}-value is found to be 9.09, at an orbital period of 8.03 days, which is above our detection threshold value of 6.
 In the lower panel of Figure~\ref{fig:groundtls} the fitted model to this 8.03 day signal is not a convincing transit, the signal most likely being a remnant of the often integer day observation cadences. 

A \ac{tls} search was performed on the TESS light curve, over an orbital period grid from 1.5 to 6 days. %
This yielded an \ac{tls} spectrum with lower powers, the maximum SDE value reached being just over 4.
With this, no significant periodicities are present in the TESS data.

\subsection{TLS Sensitivity}

\begin{figure}[htb]
    \centering
    \includegraphics[width = \hsize]{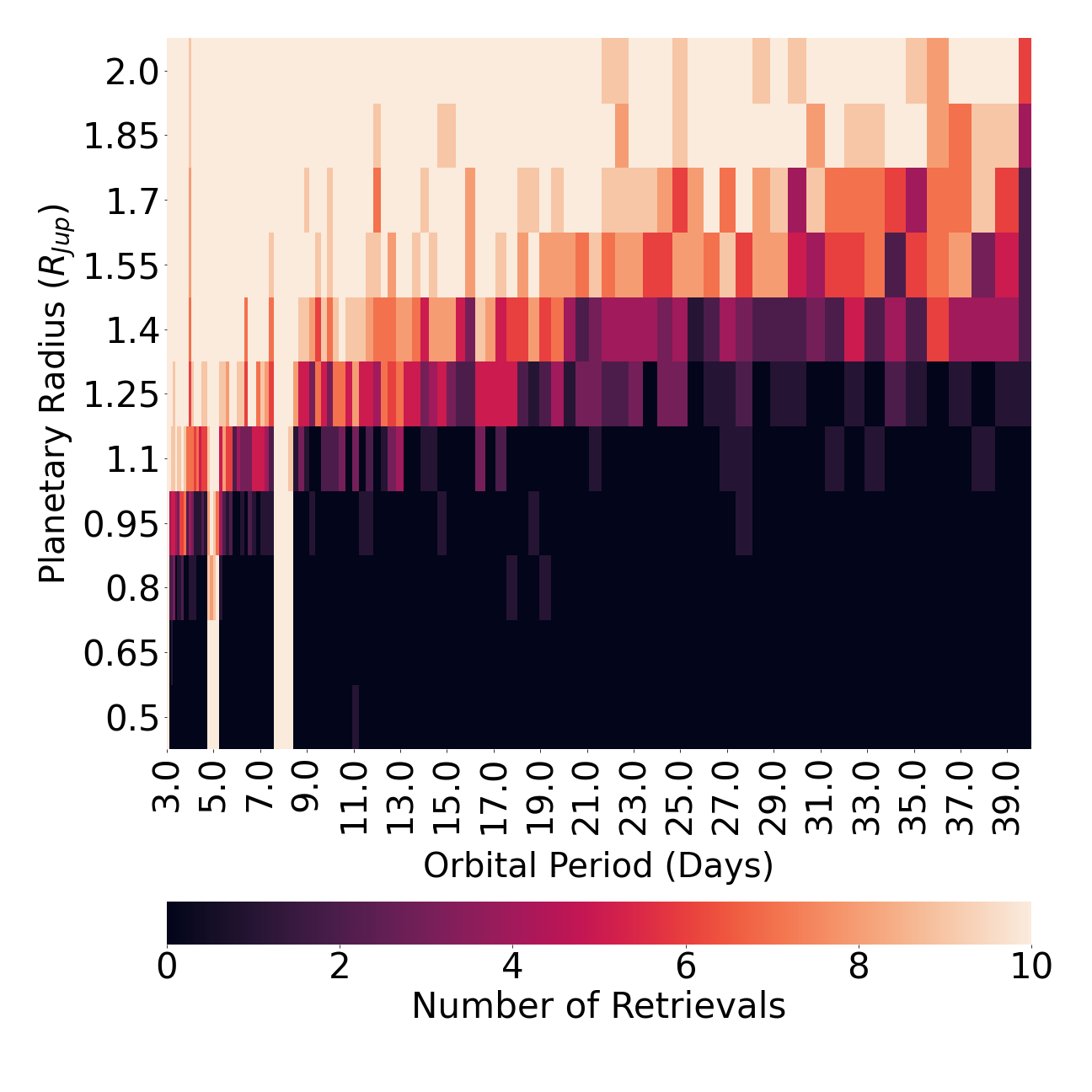}
    \caption{Sensitivity of the \ac{tls} algorithm in the J1407 system for generated transits with different radii and orbital periods injected in the real data. Columns at the right hand side of the plot are broader than ones on the left hand side, due to the logarithmic grid spacing. The corresponding masses were taken from the 20~Myr AMES-Dusty models \citep{Allard01}. } 
    \label{fig:tls_sensitivity_data}
\end{figure}

\begin{figure}[htb]
    \centering
    \includegraphics[width = \hsize]{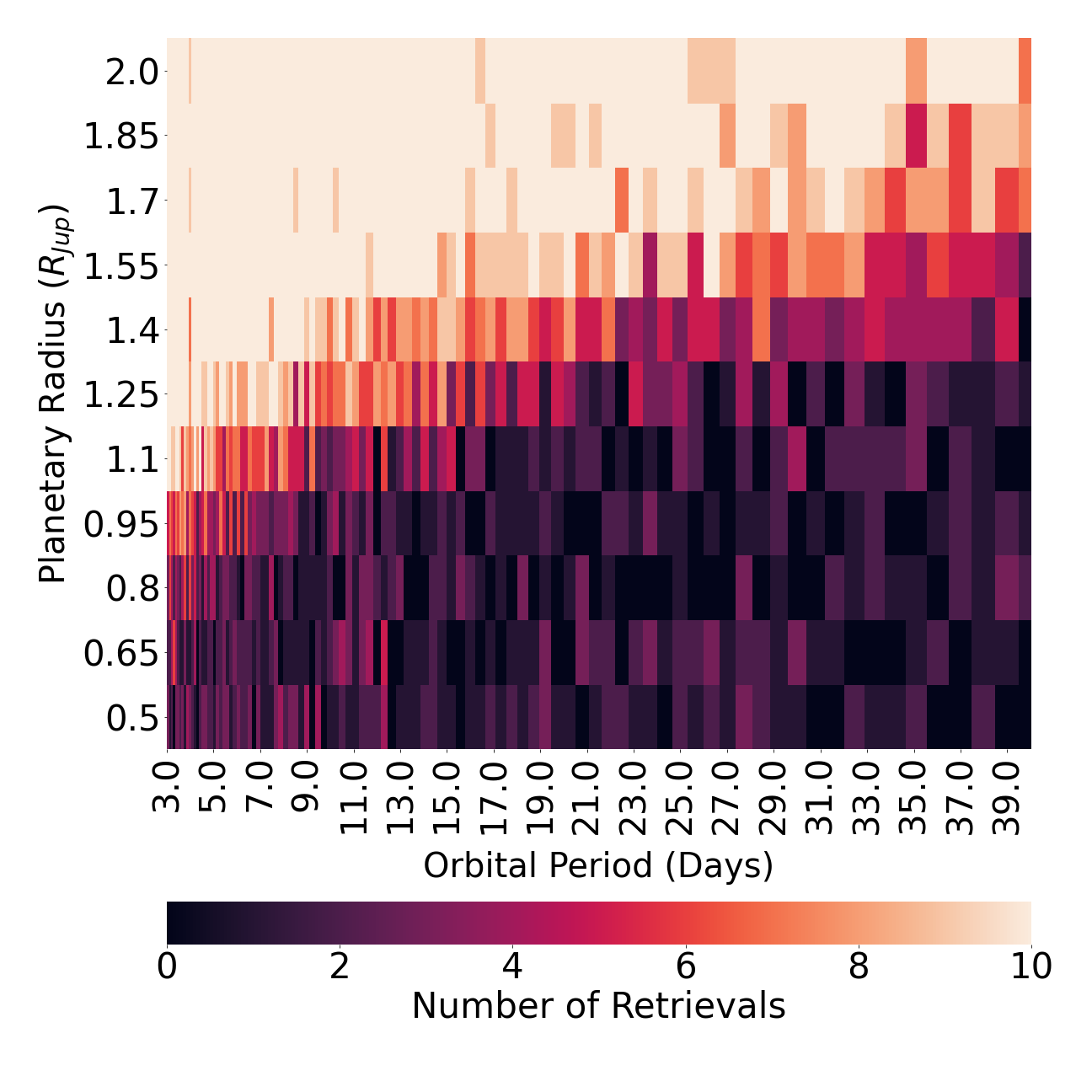}
    \caption{Sensitivity of the \ac{tls} algorithm for injected planets where the photometry is generated with white noise and the same observing epochs as the J1407 data. The corresponding masses were estimated using the 20~Myr AMES-Dusty models \citep{Allard01}. }
    \label{fig:tls_sensitivity_data_fake}
\end{figure}

The final results of the procedure described in Section~\ref{sec:TLS} can be seen in Figure~\ref{fig:tls_sensitivity_data} for the real data and in Figure~\ref{fig:tls_sensitivity_data_fake} for the generated data.
Both figures show a similar pattern, with transits of shorter orbital period and larger planetary radius being retrieved more often.
The vertical lines in Figure~\ref{fig:tls_sensitivity_data} with higher retrieval rates than their surroundings are most likely caused by remaining low power periodic signals in the data, which are just about able to pass our \ac{sde} threshold.
When a planet with similar periodicity to these remaining signals is injected, the algorithm then detects the remaining signal instead of the transit, seemingly always recovering the planet independent of radius.
These lines should therefore be interpreted as an artefact of the methodology, and not as periodicities for which we are more sensitive. 
The real retrieval rates for these periodicities are best estimated as equal to their surrounding periodicities.

Another clear trend is that the critical radius at which transits start to be retrieved is systematically smaller for the generated data.
This is due to the simulated data being generated with ideal Gaussian noise, while the real data has additional red noise due to the atmosphere and unmodeled systematic effects.

\section{Discussion}\label{sec:discuss}

No additional transiting objects were found in the J1407 system.
Upper limits on the radii of potentially missed planets were determined for different orbital periods: for a 3 day orbital period, the 50\% injected planet recovery fraction radius is 0.95\,\rjup{}, and for 10 days, 1.25\,\rjup{} and for 20 days no larger than 1.4\,\rjup.
As the system is still very young, any Jupiter-mass planet on a short orbit would also be young and hot, inflating potential planets to well above the detectable radius limits placed by our analysis.
Our \ac{tls} analysis therefore rules out planetary mass companions at shorter orbital periods.

The planetary occurrence rate for G-type (the type J1407 will be once it reaches the main sequence) stars like J1407 on our tested orbital period and radius grid is below 1\% \citep{Planetary_occurence_rate}.
The sensitivity plots (Figures~\ref{fig:tls_sensitivity_data} and \ref{fig:tls_sensitivity_data_fake}) show similar trends as the ones in the search for planets orbiting $\beta$ Pictoris in \cite{2018A&A...615A.145L}.
It should be noted that they exclude a Jupiter-sized (mass as well as radius) planet at the shortest orbital periods.

In \cite{BD_formation} it is argued that two distinct brown dwarf (BD) formation mechanisms exist; BDs with masses below 42.5 \mjup, are thought to be formed from a protoplanetary disc, while BDs with masses above 42.5 \mjup, would form more like stellar binaries.
With a mass of 60 to 100 \mjup, J1407\,b should be firmly in the latter group, although a large orbital period could potentially put the mass of J1407\,b around the lower limit of 20 \mjup  \citep{2015MNRAS.446..411K}.
Combined with the presumed high eccentricity of J1407\,b \citep[derived from the large measured transverse velocity and lack of eclipse detected for a circular orbit as discussed in][]{2015MNRAS.446..411K} the absence of greater than Jupiter-sized planets in the inner part of the system is another hint that J1407 was indeed formed more like a binary without the presence of any protoplanetary disk. 

\section{Conclusions}\label{sec:conclusion}

We model and remove the dominant periodic component from the light curve of the star J1407, and perform a search for transits of substellar companions and examine the light curve for any other anomalous features.
We detected and characterised the rotational period cycle for J1407 and it was found to be consistent for a young K-type star, using data from several ground-based telescopes with a combined baseline of 19 years.
After removing a simple model for the stellar activity, we searched the activity subtracted data, as well as 25 days of short cadence TESS data, for exoplanet transits.
No plausible transit signal was found, allowing us to constrain the presence of transiting planets within the sampled period-radius space to companions no larger than about 1.25 \,\rjup{} with periods of less than 40 days.
Further monitoring of the J1407 system continues, in anticipation of the next eclipse of J1407b.

\begin{acknowledgements}
We thank the referee for a careful reading of our manuscript - their suggestions have improved the paper.
This research has used the SIMBAD database, operated at CDS, Strasbourg, France \citep{wenger2000}.
This work has used data from the European Space Agency (ESA) mission {\it Gaia} (\url{https://www.cosmos.esa.int/gaia}), processed by the {\it Gaia} Data Processing and Analysis Consortium (DPAC, \url{https://www.cosmos.esa.int/web/gaia/dpac/consortium}).
Funding for the DPAC has been provided by national institutions, in particular the institutions participating in the {\it Gaia} Multilateral Agreement.
To achieve the scientific results presented in this article we made use of the \emph{Python} programming language\footnote{Python Software Foundation, \url{https://www.python.org/}}, especially the \emph{SciPy} \citep{virtanen2020}, \emph{NumPy} \citep{numpy}, \emph{Matplotlib} \citep{Matplotlib}, \emph{emcee} \citep{foreman-mackey2013}, and \emph{astropy} \citep{astropy_1,astropy_2} packages.
We acknowledge with thanks the variable star observations from the AAVSO International Database contributed by observers worldwide and used in this research.
We thank the Las Cumbres Observatory and its staff for its continuing support of the ASAS-SN project.
We also thank the Ohio State University College of Arts and Sciences Technology Services for helping us set up and maintain the ASAS-SN variable stars and photometry databases.
ASAS-SN is supported by the Gordon and Betty Moore Foundation through grant GBMF5490 to the Ohio State University and NSF grant AST-1515927.
Development of ASAS-SN has been supported by NSF grant AST-0908816, the Mt. Cuba Astronomical Foundation, the Center for Cosmology and AstroParticle Physics at the Ohio State University, the Chinese Academy of Sciences South America Center for Astronomy (CASSACA), the Villum Foundation, and George Skestos.
Early work on KELT-North was supported by NASA Grant NNG04GO70G.
Work on KELT-North was partially supported by NSF CAREER Grant AST-1056524 to S. Gaudi.
Work on KELT-North received support from the Vanderbilt Office of the Provost through the Vanderbilt Initiative in Data-intensive Astrophysics.
Part of this research was carried out in part at the Jet Propulsion Laboratory, California Institute of Technology, under a contract with the National Aeronautics and Space Administration (80NM0018D0004).

\end{acknowledgements}

\bibliography{references}

\begin{thebibliography}{43}
\expandafter\ifx\csname natexlab\endcsname\relax\def\natexlab#1{#1}\fi

\bibitem[{{Allard} {et~al.}(2001){Allard}, {Hauschildt}, {Alexander},
  {Tamanai}, \& {Schweitzer}}]{Allard01}
{Allard}, F., {Hauschildt}, P.~H., {Alexander}, D.~R., {Tamanai}, A., \&
  {Schweitzer}, A. 2001, \apj, 556, 357

\bibitem[{{Astropy Collaboration} {et~al.}(2018){Astropy Collaboration},
  {Price-Whelan}, {Sip{\H o}cz}, {G{\"u}nther}, {Lim}, {Crawford}, {Conseil},
  {Shupe}, {Craig}, {Dencheva}, {Ginsburg}, {VanderPlas}, {Bradley},
  {P{\'e}rez-Su{\'a}rez}, {de Val-Borro}, {Aldcroft}, {Cruz}, {Robitaille},
  {Tollerud}, {Ardelean}, {Babej}, {Bach}, {Bachetti}, {Bakanov}, {Bamford},
  {Barentsen}, {Barmby}, {Baumbach}, {Berry}, {Biscani}, {Boquien}, {Bostroem},
  {Bouma}, {Brammer}, {Bray}, {Breytenbach}, {Buddelmeijer}, {Burke},
  {Calderone}, {Cano Rodr{\'{\i}}guez}, {Cara}, {Cardoso}, {Cheedella},
  {Copin}, {Corrales}, {Crichton}, {D'Avella}, {Deil}, {Depagne}, {Dietrich},
  {Donath}, {Droettboom}, {Earl}, {Erben}, {Fabbro}, {Ferreira}, {Finethy},
  {Fox}, {Garrison}, {Gibbons}, {Goldstein}, {Gommers}, {Greco}, {Greenfield},
  {Groener}, {Grollier}, {Hagen}, {Hirst}, {Homeier}, {Horton}, {Hosseinzadeh},
  {Hu}, {Hunkeler}, {Ivezi{\'c}}, {Jain}, {Jenness}, {Kanarek}, {Kendrew},
  {Kern}, {Kerzendorf}, {Khvalko}, {King}, {Kirkby}, {Kulkarni}, {Kumar},
  {Lee}, {Lenz}, {Littlefair}, {Ma}, {Macleod}, {Mastropietro}, {McCully},
  {Montagnac}, {Morris}, {Mueller}, {Mumford}, {Muna}, {Murphy}, {Nelson},
  {Nguyen}, {Ninan}, {N{\"o}the}, {Ogaz}, {Oh}, {Parejko}, {Parley}, {Pascual},
  {Patil}, {Patil}, {Plunkett}, {Prochaska}, {Rastogi}, {Reddy Janga},
  {Sabater}, {Sakurikar}, {Seifert}, {Sherbert}, {Sherwood-Taylor}, {Shih},
  {Sick}, {Silbiger}, {Singanamalla}, {Singer}, {Sladen}, {Sooley},
  {Sornarajah}, {Streicher}, {Teuben}, {Thomas}, {Tremblay}, {Turner},
  {Terr{\'o}n}, {van Kerkwijk}, {de la Vega}, {Watkins}, {Weaver}, {Whitmore},
  {Woillez}, {Zabalza}, \& {Astropy Contributors}}]{astropy_2}
{Astropy Collaboration}, {Price-Whelan}, A.~M., {Sip{\H o}cz}, B.~M., {et~al.}
  2018, \aj, 156, 123

\bibitem[{{Astropy Collaboration} {et~al.}(2013){Astropy Collaboration},
  {Robitaille}, {Tollerud}, {Greenfield}, {Droettboom}, {Bray}, {Aldcroft},
  {Davis}, {Ginsburg}, {Price-Whelan}, {Kerzendorf}, {Conley}, {Crighton},
  {Barbary}, {Muna}, {Ferguson}, {Grollier}, {Parikh}, {Nair}, {Unther},
  {Deil}, {Woillez}, {Conseil}, {Kramer}, {Turner}, {Singer}, {Fox}, {Weaver},
  {Zabalza}, {Edwards}, {Azalee Bostroem}, {Burke}, {Casey}, {Crawford},
  {Dencheva}, {Ely}, {Jenness}, {Labrie}, {Lim}, {Pierfederici}, {Pontzen},
  {Ptak}, {Refsdal}, {Servillat}, \& {Streicher}}]{astropy_1}
{Astropy Collaboration}, {Robitaille}, T.~P., {Tollerud}, E.~J., {et~al.} 2013,
  \aap, 558, A33

\bibitem[{{Braga-Ribas} {et~al.}(2014){Braga-Ribas}, {Sicardy}, {Ortiz},
  {Snodgrass}, {Roques}, {Vieira-Martins}, {Camargo}, {Assafin}, {Duffard},
  {Jehin}, {Pollock}, {Leiva}, {Emilio}, {Machado}, {Colazo}, {Lellouch},
  {Skottfelt}, {Gillon}, {Ligier}, {Maquet}, {Benedetti-Rossi}, {Gomes},
  {Kervella}, {Monteiro}, {Sfair}, {El Moutamid}, {Tancredi}, {Spagnotto},
  {Maury}, {Morales}, {Gil-Hutton}, {Roland}, {Ceretta}, {Gu}, {Wang},
  {Harps{\o}e}, {Rabus}, {Manfroid}, {Opitom}, {Vanzi}, {Mehret}, {Lorenzini},
  {Schneiter}, {Melia}, {Lecacheux}, {Colas}, {Vachier}, {Widemann},
  {Almenares}, {Sandness}, {Char}, {Perez}, {Lemos}, {Martinez},
  {J{\o}rgensen}, {Dominik}, {Roig}, {Reichart}, {Lacluyze}, {Haislip},
  {Ivarsen}, {Moore}, {Frank}, \& {Lambas}}]{2014Natur.508...72B}
{Braga-Ribas}, F., {Sicardy}, B., {Ortiz}, J.~L., {et~al.} 2014, \nat, 508, 72

\bibitem[{{Charnoz} {et~al.}(2018){Charnoz}, {Crida}, \&
  {Hyodo}}]{2018haex.bookE..54C}
{Charnoz}, S., {Crida}, A., \& {Hyodo}, R. 2018, {Rings in the Solar System: A
  Short Review} (Springer International Publishing AG), 54

\bibitem[{{Claret} \& {Bloemen}(2011)}]{2011yCat..35290075C}
{Claret}, A. \& {Bloemen}, S. 2011, VizieR Online Data Catalog, J/A+A/529/A75

\bibitem[{{Feinstein} {et~al.}(2019){Feinstein}, {Montet}, {Foreman-Mackey},
  {Bedell}, {Saunders}, {Bean}, {Christiansen}, {Hedges}, {Luger}, {Scolnic},
  \& {Cardoso}}]{2019PASP..131i4502F}
{Feinstein}, A.~D., {Montet}, B.~T., {Foreman-Mackey}, D., {et~al.} 2019,
  \pasp, 131, 094502

\bibitem[{{Foreman-Mackey} {et~al.}(2013{\natexlab{a}}){Foreman-Mackey},
  {Hogg}, {Lang}, \& {Goodman}}]{emcee}
{Foreman-Mackey}, D., {Hogg}, D.~W., {Lang}, D., \& {Goodman}, J.
  2013{\natexlab{a}}, \pasp, 125, 306

\bibitem[{{Foreman-Mackey} {et~al.}(2013{\natexlab{b}}){Foreman-Mackey},
  {Hogg}, {Lang}, \& {Goodman}}]{foreman-mackey2013}
{Foreman-Mackey}, D., {Hogg}, D.~W., {Lang}, D., \& {Goodman}, J.
  2013{\natexlab{b}}, \pasp, 125, 306

\bibitem[{{Guinan} \& {Dewarf}(2002)}]{2002ASPC..279..121G}
{Guinan}, E.~F. \& {Dewarf}, L.~E. 2002, Astronomical Society of the Pacific
  Conference Series, Vol. 279, {Toward Solving the Mysteries of the Exotic
  Eclipsing Binary {\ensuremath{\in}} Aurigae: Two Thousand years of
  Observations and Future Possibilities} (PASP), 121

\bibitem[{{Hambsch}(2012)}]{2012JAVSO..40.1003H}
{Hambsch}, F.~J. 2012, Journal of the American Association of Variable Star
  Observers (JAAVSO), 40, 1003

\bibitem[{{Hathaway}(2015)}]{Hathaway_2015}
{Hathaway}, D.~H. 2015, Living Reviews in Solar Physics, 12, 4

\bibitem[{{Hippke} \& {Heller}(2019)}]{Hippke_2019}
{Hippke}, M. \& {Heller}, R. 2019, Astronomy and Astrophysics, 623, A39

\bibitem[{{Hunter}(2007)}]{Matplotlib}
{Hunter}, J.~D. 2007, Computing in Science and Engineering, 9, 90

\bibitem[{{Kenworthy} {et~al.}(2015{\natexlab{a}}){Kenworthy}, {Lacour},
  {Kraus}, {Triaud}, {Mamajek}, {Scott}, {S{\'e}gransan}, {Ireland}, {Hambsch},
  {Reichart}, {Haislip}, {LaCluyze}, {Moore}, \& {Frank}}]{Kenworthy15}
{Kenworthy}, M.~A., {Lacour}, S., {Kraus}, A., {et~al.} 2015{\natexlab{a}},
  \mnras, 446, 411

\bibitem[{{Kenworthy} {et~al.}(2015{\natexlab{b}}){Kenworthy}, {Lacour},
  {Kraus}, {Triaud}, {Mamajek}, {Scott}, {S{\'e}gransan}, {Ireland}, {Hambsch},
  {Reichart}, {Haislip}, {LaCluyze}, {Moore}, \& {Frank}}]{2015MNRAS.446..411K}
{Kenworthy}, M.~A., {Lacour}, S., {Kraus}, A., {et~al.} 2015{\natexlab{b}},
  \mnras, 446, 411

\bibitem[{{Kenworthy} \& {Mamajek}(2015)}]{Kenworthy_2015}
{Kenworthy}, M.~A. \& {Mamajek}, E.~E. 2015, Astrophysical Journal, 800, 126

\bibitem[{{Kochanek} {et~al.}(2017){Kochanek}, {Shappee}, {Stanek}, {Holoien},
  {Thompson}, {Prieto}, {Dong}, {Shields}, {Will}, {Britt}, {Perzanowski}, \&
  {Pojma{\'n}ski}}]{2017PASP..129j4502K}
{Kochanek}, C.~S., {Shappee}, B.~J., {Stanek}, K.~Z., {et~al.} 2017,
  Publications of the ASP, 129, 104502

\bibitem[{{Kov{\'a}cs} {et~al.}(2002){Kov{\'a}cs}, {Zucker}, \&
  {Mazeh}}]{Kov_cs_2002}
{Kov{\'a}cs}, G., {Zucker}, S., \& {Mazeh}, T. 2002, Astronomy and
  Astrophysics, 391, 369

\bibitem[{{Kreidberg}(2015)}]{Kreidberg_2015}
{Kreidberg}, L. 2015, Publications of the ASP, 127, 1161

\bibitem[{{Kunimoto} \& {Matthews}(2020)}]{Planetary_occurence_rate}
{Kunimoto}, M. \& {Matthews}, J.~M. 2020, Astronomical Journal, 159, 248

\bibitem[{{Lipunov} {et~al.}(2016){Lipunov}, {Gorbovskoy}, {Afanasiev},
  {Tatarnikova}, {Denisenko}, {Makarov}, {Tiurina}, {Krushinsky}, {Vinokurov},
  {Balanutsa}, {Kuznetsov}, {Gress}, {Sergienko}, {Yurkov}, {Gabovich},
  {Tlatov}, {Senik}, {Vladimirov}, \& {Popova}}]{2016A&A...588A..90L}
{Lipunov}, V., {Gorbovskoy}, E., {Afanasiev}, V., {et~al.} 2016, \aap, 588, A90

\bibitem[{{Lous} {et~al.}(2018){Lous}, {Weenk}, {Kenworthy}, {Zwintz}, \&
  {Kuschnig}}]{2018A&A...615A.145L}
{Lous}, M.~M., {Weenk}, E., {Kenworthy}, M.~A., {Zwintz}, K., \& {Kuschnig}, R.
  2018, Astronomy and Astrophysics,, 615, A145

\bibitem[{{Ma} \& {Ge}(2014)}]{BD_formation}
{Ma}, B. \& {Ge}, J. 2014, Monthly Notices of the RAS, 439, 2781

\bibitem[{{Mamajek} {et~al.}(2012){Mamajek}, {Quillen}, {Pecaut}, {Moolekamp},
  {Scott}, {Kenworthy}, {Collier Cameron}, \& {Parley}}]{2012AJ....143...72M}
{Mamajek}, E.~E., {Quillen}, A.~C., {Pecaut}, M.~J., {et~al.} 2012,
  Astronomical Journal,, 143, 72

\bibitem[{{Mentel} {et~al.}(2018){Mentel}, {Kenworthy}, {Cameron}, {Scott},
  {Mellon}, {Hudec}, {Birkby}, {Mamajek}, {Schrimpf}, {Reichart}, {Haislip},
  {Kouprianov}, {Hambsch}, {Tan}, {Hills}, {Grindlay}, {Rodriguez}, {Lund}, \&
  {Kuhn}}]{2018A&A...619A.157M}
{Mentel}, R.~T., {Kenworthy}, M.~A., {Cameron}, D.~A., {et~al.} 2018, Astronomy
  and Astrophysics,, 619, A157

\bibitem[{{Mikolajewski} \& {Graczyk}(1999)}]{10.1046/j.1365-8711.1999.02257.x}
{Mikolajewski}, M. \& {Graczyk}, D. 1999, Monthly Notices of the RAS, 303, 521

\bibitem[{{Newville} {et~al.}(2014){Newville}, {Stensitzki}, {Allen}, \&
  {Ingargiola}}]{lmfit}
{Newville}, M., {Stensitzki}, T., {Allen}, D.~B., \& {Ingargiola}, A. 2014,
  {LMFIT: Non-Linear Least-Square Minimization and Curve-Fitting for Python}

\bibitem[{Oliphant(2006)}]{numpy}
Oliphant, T.~E. 2006, A guide to NumPy, Vol.~1 (Trelgol Publishing USA)

\bibitem[{{Pepper} {et~al.}(2012){Pepper}, {Kuhn}, {Siverd}, {James}, \&
  {Stassun}}]{Pepper12}
{Pepper}, J., {Kuhn}, R.~B., {Siverd}, R., {James}, D., \& {Stassun}, K. 2012,
  \pasp, 124, 230

\bibitem[{{Pepper} {et~al.}(2007){Pepper}, {Pogge}, {DePoy}, {Marshall},
  {Stanek}, {Stutz}, {Poindexter}, {Siverd}, {O'Brien}, {Trueblood}, \&
  {Trueblood}}]{2007PASP..119..923P}
{Pepper}, J., {Pogge}, R.~W., {DePoy}, D.~L., {et~al.} 2007, Publications of
  the ASP, 119, 923

\bibitem[{{Pizzolato} {et~al.}(2003){Pizzolato}, {Maggio}, {Micela},
  {Sciortino}, \& {Ventura}}]{Pizzolato03}
{Pizzolato}, N., {Maggio}, A., {Micela}, G., {Sciortino}, S., \& {Ventura}, P.
  2003, \aap, 397, 147

\bibitem[{{Pojmanski}(1997)}]{1997AcA....47..467P}
{Pojmanski}, G. 1997, Acta Astronomica, 47, 467

\bibitem[{{Reichart} {et~al.}(2005){Reichart}, {Nysewander}, {Moran},
  {Bartelme}, {Bayliss}, {Foster}, {Clemens}, {Price}, {Evans}, {Salmonson},
  {Trammell}, {Carney}, {Keohane}, \& {Gotwals}}]{2005NCimC..28..767R}
{Reichart}, D., {Nysewander}, M., {Moran}, J., {et~al.} 2005, Nuovo Cimento C
  Geophysics Space Physics C, 28, 767

\bibitem[{{Ricker} {et~al.}(2015){Ricker}, {Winn}, {Vanderspek}, {Latham},
  {Bakos}, {Bean}, {Berta-Thompson}, {Brown}, {Buchhave}, {Butler}, {Butler},
  {Chaplin}, {Charbonneau}, {Christensen-Dalsgaard}, {Clampin}, {Deming},
  {Doty}, {De Lee}, {Dressing}, {Dunham}, {Endl}, {Fressin}, {Ge}, {Henning},
  {Holman}, {Howard}, {Ida}, {Jenkins}, {Jernigan}, {Johnson}, {Kaltenegger},
  {Kawai}, {Kjeldsen}, {Laughlin}, {Levine}, {Lin}, {Lissauer}, {MacQueen},
  {Marcy}, {McCullough}, {Morton}, {Narita}, {Paegert}, {Palle}, {Pepe},
  {Pepper}, {Quirrenbach}, {Rinehart}, {Sasselov}, {Sato}, {Seager},
  {Sozzetti}, {Stassun}, {Sullivan}, {Szentgyorgyi}, {Torres}, {Udry}, \&
  {Villasenor}}]{2015JATIS...1a4003R}
{Ricker}, G.~R., {Winn}, J.~N., {Vanderspek}, R., {et~al.} 2015, Journal of
  Astronomical Telescopes, Instruments, and Systems, 1, 014003

\bibitem[{{Rieder} \& {Kenworthy}(2016)}]{Rieder_2016}
{Rieder}, S. \& {Kenworthy}, M.~A. 2016, Astronomy and Astrophysics, 596, A9

\bibitem[{{Rodriguez} {et~al.}(2016){Rodriguez}, {Stassun}, {Lund}, {Siverd},
  {Pepper}, {Tang}, {Kafka}, {Gaudi}, {Conroy}, {Beatty}, {Stevens}, {Shappee},
  \& {Kochanek}}]{2016AJ....151..123R}
{Rodriguez}, J.~E., {Stassun}, K.~G., {Lund}, M.~B., {et~al.} 2016, \aj, 151,
  123

\bibitem[{{Saar} \& {Brandenburg}(1999)}]{cycle_periods_saar_1999}
{Saar}, S.~H. \& {Brandenburg}, A. 1999, Astrophysical Journal, 524, 295

\bibitem[{{Tiscareno}(2013)}]{2013pss3.book..309T}
{Tiscareno}, M.~S. 2013, {Planetary Rings} (Springer Science+Business Media
  Dordrecht), 309

\bibitem[{{Virtanen} {et~al.}(2020){Virtanen}, {Gommers}, {Oliphant},
  {Haberland}, {Reddy}, {Cournapeau}, {Burovski}, {Peterson}, {Weckesser},
  {Bright}, {van der Walt}, {Brett}, {Wilson}, {Millman}, {Mayorov}, {Nelson},
  {Jones}, {Kern}, {Larson}, {Carey}, {Polat}, {Feng}, {Moore}, {Vand erPlas},
  {Laxalde}, {Perktold}, {Cimrman}, {Henriksen}, {Quintero}, {Harris},
  {Archibald}, {Ribeiro}, {Pedregosa}, {van Mulbregt}, \& {SciPy 1. 0
  Contributors}}]{virtanen2020}
{Virtanen}, P., {Gommers}, R., {Oliphant}, T.~E., {et~al.} 2020, Nature
  Methods, 17, 261

\bibitem[{{Wenger} {et~al.}(2000){Wenger}, {Ochsenbein}, {Egret}, {Dubois},
  {Bonnarel}, {Borde}, {Genova}, {Jasniewicz}, {Lalo{\"e}}, {Lesteven}, \&
  {Monier}}]{wenger2000}
{Wenger}, M., {Ochsenbein}, F., {Egret}, D., {et~al.} 2000, \aaps, 143, 9

\bibitem[{{Willamo} {et~al.}(2020){Willamo}, {Hackman}, {Lehtinen},
  {K{\"a}pyl{\"a}}, {Olspert}, {Viviani}, \& {Warnecke}}]{2020A&A...638A..69W}
{Willamo}, T., {Hackman}, T., {Lehtinen}, J.~J., {et~al.} 2020, \aap, 638, A69

\bibitem[{{Zanazzi} \& {Lai}(2017)}]{Zanezzi}
{Zanazzi}, J.~J. \& {Lai}, D. 2017, Monthly Notices of the RAS, 464, 3945

\end{thebibliography}

\end{document}